\documentclass[12pt,preprint]{aastex}
\usepackage{graphicx}

\newcommand{\zsolar}{Z$_{\odot}$}
\newcommand{\msolar}{M$_{\odot}$}
\newcommand{\ud}{\mathrm{d}}

\newcommand{\lsolar}{L$_{\odot}$}

\newcommand{\mic}{\mbox{$\mu$m}}
\newcommand{\sbs}{\mbox{SBS\,0335-052}}
\newcommand{\hen}{\mbox{He\,2-10}}
\newcommand{\zw}{\mbox{I\,Zw\,18}}
\newcommand{\hii}{\mbox{H{\sc ii}}}

\shorttitle{The embedded super star cluster of \sbs}
\shortauthors{Plante \& Sauvage}

\begin{document}

\title{The Embedded Super Star Cluster of \sbs \footnote{This paper is
based: (1)  on
data obtained with ISO, an ESA project with instruments funded by
the ESA member states (especially the PI countries: France, Germany,
the Netherlands, and the United Kingdom) with the participation of
ISAS and NASA, (2) on observations obtained at the Gemini Observatory, which is
operated by the Association of Universities for Research in
Astronomy, Inc., under a cooperative agreement with the NSF on
behalf of the Gemini partnership: the National Science Foundation
(United States), the Particle Physics and Astronomy Research
Council (United Kingdom), the National Research Council (Canada),
CONICYT (Chile), the Australian Research Council (Australia), CNPq
(Brazil) and CONICET (Argentina), using the mid-infrared camera OSCIR,
developed by the University of Florida with support from the National
Aeronautics and Space Administration, and operated jointly by Gemini
and the University of Florida Infrared Astrophysics Group.}}

\author{St\'ephanie Plante}
\affil{D\'epartement de physique, de g\'enie
physique et d'optique, Universit\'e Laval and Observatoire du Mont
M\'egantic, Qu\'ebec, QC, Canada, G1K 7P4 }
\email{splante@phy.ulaval.ca}

\and

\author{Marc Sauvage}
\affil{DAPNIA/Service d'Astrophysique,
CEA/Saclay, 91191 Gif-sur-Yvette Cedex, France}
\email{msauvage@cea.fr}

\begin{abstract}
    We analyze the infrared (6-100\,\mic) spectral energy distribution of
    the blue compact dwarf and metal-poor (Z=\zsolar/41) galaxy \sbs.
    With the help of DUSTY \citep{Ive99}, a program that solves the
    radiation transfer equations in a spherical environment, we evaluate
    that the infrared (IR) emission of \sbs~is produced by an embedded
    super-star cluster (SSC) hidden under 10$^5$ \msolar~of dust, causing
    30 mag of visual extinction.  This implies that one cannot detect any
    stellar emission from the 2$\times10^6$ \msolar\ stellar cluster even
    at near-infrared (NIR) wavelengths.  The derived grain size
    distribution departs markedly from the widely accepted size
    distribution inferred for dust in our galaxy (the so-called MRN
    distribution, \citet{Mat77}), but resembles what is seen around AGNs,
    namely an absence of PAH and smaller grains, and grains that grow to
    larger sizes (around 1 \micron).  The fact that a significant amount
    of dust is present in such a low-metallicity galaxy, hiding from UV
    and optical view most of the star formation activity in the galaxy,
    and that the dust size distribution cannot be reproduced by a
    standard galactic law, should be borne in mind when interpreting
the spectrum of
    primeval galaxies.
\end{abstract}

\section{Introduction}
\label{introduction}
The question of how the energy radiated by a very young burst of star
formation is redistributed in the electromagnetic spectrum by the
neighboring ISM is one with far-reaching implications.  Indeed, as it
is generally assumed that the formation of galaxies should be
signalled by violent bursts of star formation (see e.g. the reviews by
\citet{Sil01} or \citet{Ell98}, and references therein), the answer to
this question can help defining the best observing strategy to study
primeval galaxies.  For the most massive objects,
it is generally assumed that star formation proceeds as observed in
Ultra-Luminous InfraRed Galaxies (ULIRGs, see the review by
\citet{San96}).  In these systems, we know that most of the energy
emerges in the infrared, and this would tend to invalidate any result
derived from optical-UV surveys of the distant Universe.  Yet studies
on the local starburst galaxy population appear to indicate a
correlation between the total infrared luminosity and the extinction
as measured by the slope of the UV continuum \citep{Meu95}.  Such a
correlation, along with the establishment of an effective attenuation
curve \citep{Cal94}, offers the hope to address the
question of galaxy formation with optical-UV instruments, thus
circumventing an important problem of most current infrared and
submillimeter instruments: their lower spatial resolution that makes
the identification of counterparts and subsequent determination of
redshifts problematic.

However a number of relatively recent discoveries on the
properties of starburst galaxies and ULIRGs cast some doubt on the
potential of this UV-IR/Submm relation and on the physical meaning
of an attenuation curve for getting at the intrinsic UV luminosity
of a starburst galaxy. Recent high-spatial resolution MIR
instruments have revealed the existence of very bright super
star-clusters (clusters containing a few thousand O stars,
hereafter SSCs) that are nearly or absolutely absent from visible
images, e.g. the deeply
buried SSC found in the Antennae galaxy \citep{mvc98}.
This object produces about 20\% of the total MIR emission of
the whole galaxy and was shown by \citet{gil00} to be a very young
($\sim$ 4\,Myr) SSC containing 1.6$\times10^{7}$\,\msolar\ of stars
embedded in an $A_{V}=10$ cloud of dust. This is no longer an
isolated case: the Wolf-Rayet dwarf galaxy \hen\ is an even more
impressive example of the buried SSC phenomenology. \citet{Kob99}
showed that \hen\ contains extremely compact radio sources whose
spectrum is optically thick at 5\,GHz, which are
interpreted as ultra-dense \hii\ regions created by dust-embedded SSCs each
with $\sim$750\,O7{\sc V} stars. Gemini/OSCIR high resolution
MIR observations by \citet{Vac02} showed that the radio SSCs are
exactly coincident with the MIR emitting regions observed
previously by \citet{Stl97}; the SSCs generate {\em
almost all} of the MIR luminosity of the galaxy, and
there is {\em no overlap} between the MIR
emitting regions and those detected in the K band. This is also
true with the L and M bands \citep{Sau02} implying a very high
optical depth along the line of sight toward the SSCs. Another case
where the infrared emission arises from a dust-embedded SSC with
no optical counterpart is the dwarf galaxy NGC\,5253 \citep{Tur00,
Gor01}.

Recent observations have shown that dust is present even in
the most metal-deficient objects in amounts large enough to affect our
ability to observe the star-formation process, namely \zw\ and \sbs.
In \zw, still the most metal-poor galaxy known at Z=\zsolar/50,
the analysis of the H$\alpha$/H$\beta$ ratio by \citet{Can02} indicates
patches of dust inside the \hii\ regions that lead to
$A_{V}=0.5$\,mag in some places.

\sbs, at \zsolar/41, for which we are presenting new data, has the
highest star formation rate of the two.  \citet{Thu97}, based on HST
images, argue that this galaxy is probably undergoing its first burst
of star formation (but see \citet{Ost01}, and consider that the aim of
this paper is {\em not} to discuss whether or not \sbs\ undergoes its
first burst of star formation, but rather to show that the {\em
current} burst properties can shed light on phenomena possibly
occurring in primeval galaxies.  In other words, \sbs\ is considered
in this work as a laboratory to study primeval galaxies, but not as a
primeval galaxy itself).  In HST images, young stars appears
concentrated in 6 SSCs, each of them not older than 25 Myr and all
located within a region smaller than 526 pc\footnote{With H$_o$ = 75
km s$^{-1}$ Mpc$^{-1}$.  At this distance, 1${\arcsec}$ is 263 pc.}.
In the NIR, the emission originates mostly from a region coincident
with two of these SSCs (the ground-based NIR image does not allow to
precisely attribute the emission to the HST-detected SSCs).  The NIR
spectrum indicates stellar populations younger than 5\,Myr
\citep{Van00}.  The picture gets more complex when the MIR properties
are considered as well: the galaxy is very bright in the MIR and its
global MIR spectrum is quite unusual (\cite{Thu99}, hereafter Paper~I).
First, it is lacking the Unidentified Infrared Bands (UIB) commonly
attributed to Polycyclic Aromatic Hydrocarbons or PAHs, \citep{Leg84,
All85}.  This is generally indicative of dust exposed to a strong
radiation field that either destroys the UIB carriers or swamps their
emission in that of the very small grains.  Second it shows a marked
silicate absorption band at 9.7\,\mic, very unusual at the galaxy
scale and indicative of a large dust column density, unexpected in
such a low metallicity galaxy.  This peculiar spectrum led to the
hypothesis that the MIR emission originates from a dust-enshrouded
SSC. Subsequent ground-based observations by \citet{Dale01} showed the MIR
emission to be almost a point source coincident with the NIR emitting
region; Contrary to Paper~I, these authors argued against the buried
SSC case for \sbs.  Thus whether \sbs\ contains one or more deeply
buried SSCs remains an open question (many different $A_V$ have been
determined for the SSCs of \sbs, ranging from $A_V \sim 0.55$ based on
the Balmer decrement \citep{Izo97} to $A_V \sim 20.0$ based on MIR
spectroscopy, Paper~I), and is worth returning to.

In section~\S\ref{observations} we present new GEMINI/OSCIR
and ISOPHOT observations used in conjunction with
the ISOCAM data to reconstruct the infrared spectral energy distribution
of the galaxy. In section~\S\ref{model} we define and justify our assumptions
regarding the modelling of radiation transfer in \sbs. Our results
are presented in section~\S\ref{results}, and their implications are
discussed in section~\S\ref{discussion}.

\section{Observations}
\label{observations}

The data used to model the infrared spectral energy distribution of
\sbs\ come from three sources.  The MIR ISOCAM data were presented in
Paper~I. The ISOPHOT 60-100\,\mic\ data were obtained in the same
program as Paper~I but their analysis deferred until a full modelling of
the SED was possible.  The GEMINI MIR data were obtained as a
follow-up to clarify the issues developed in Paper~I.

\subsection{ISOPHOT observations}
\label{phot-obs}
The ISOPHOT \citep{Lemke96} data were obtained on revolution 845,
using the observation template PHT22, which consists in a small raster
around the target.  \sbs\ was observed with the C100 detector, in
three relatively broad-band filters, namely the 50\,\mic\ filter
($\lambda_{ref}=65\,\mic, \Delta\lambda=57.8\,\mic$), the 60\,\mic\
filter ($\lambda_{ref}=60\,\mic, \Delta\lambda=23.9\,\mic$), and the
100\,\mic\ filter (($\lambda_{ref}=100\,\mic,
\Delta\lambda=43.6\,\mic$).  As some confusion may arise regarding the
relative positioning of the 50 and 60\,\mic\ filters, we refer to the
ISOPHOT filters by their reference wavelengths, i.e. 60, 65 and
100\,\mic.  The definition of ISOPHOT bandpasses and spectral
conventions can be found in \citet{Laureijs00}, and have been used in
this paper when comparing our model SED to the observations.  All
three data sets were acquired in a similar fashion, with the
3$\times$3 pixel C100 detector performing a $3\times$3 raster around
the source position.  The raster axes were aligned with that of the
detector and the step between each of the raster point was equal to
the pixel size (43$\farcs$5) so that each pixel sees the center of the
field once during the observation (or equivalently, the center of the
field is observed by all nine pixels).  Operation of the C100 detector
consists in a series of non-destructive readouts for a given
integration time, called integration ramps, after which a destructive
readout resets the detector and a new ramp starts.  For all three
rasters, the individual integration ramps consisted in 64 readouts.
Each ramp lasted 2\,s.  For the 60 and 100\,\mic\ filters we took 32
integration ramps per raster position, while this was doubled to 64
ramps for the 65\,\mic\ filter.

To analyze the data, we combined the standard PHT reduction steps with
a series of algorithms designed to take into account the fact that the
source is extremely faint, and very likely point-like (it is not
resolved by ISOCAM, and barely by 8\,m telescopes on the ground, see
later and \citet{Dale01}).  The standard data reduction steps were
performed with PIA\,7.0\footnote{PIA is a joint development by the ESA
Astrophysics Division and the ISOPHOT consortium, with the
collaboration of the Infrared Analysis and Processing Center (IPAC)
and the Instituto de Astrofisica de Canarias (IAC).}.  We will not
detail here the entire process but rather show at which points we have
branched personally developed algorithms.  At the ramp stage, we found
that the deglitching methods available in PIA\,7.0 were not as robust
and discriminating as was necessary, and we used instead an
adaptation of the multi-resolution deglitching method designed for ISOCAM
data \citep{Starck99}.  At this stage a very
small percentage ($\leq$\,1\,\%) of the readouts is discarded because
glitches typically appear to affect only one of the 64 readouts
per ramp.  However, once we computed the ramp slopes, we found that
the slope signal showed a large number of positive spikes, extremely
reminiscent of glitches, affecting up to three consecutive ramps.
Inspecting the list of discarded readouts at the ramp stage, we found
that {\em all} the slope spikes could be tied to a glitch impact
affecting a readout in the corresponding ramp or the one just
preceding (although not all glitches at the readout level lead to a
spike in the ramp).  This is very similar to glitches with ``memory
effect'' experienced in the CAM LW detector \citep{Starck99}, and
given the similarities in the underlying detector physics, we
attribute these spikes to cosmic ray impacts.  Since a large number of
ramps were obtained per raster position, a multi-resolution method again proved
extremely efficient.

Due to the lack of a physical modelling of the transient
behavior of the C100 detector, we decided not to apply any transient
correction to our data. At the current stage, this would in fact
correspond to an arbitrary choice of a correcting function.

Finally, the last stage of the reduction, the map reconstruction
was also replaced by a better suited algorithm. Indeed inspection
of the signal from individual pixels revealed that the source only
illuminates one pixel of the detector at a given time, i.e. that
it is point-like for ISOPHOT. Therefore a simpler method for
detecting and measuring the source flux is to use each pixel of
the detector as a scanner and co-add these scans (obviously taking
into account the fact that the source appears at a different
position along each scan). Assuming that the background around the
source is constant, we use the off-source sectors of the scan to
derive the flat-field of the detector. Finally, the point-spread
function profile is used to extract the source flux.

With this processing, the source is clearly detected at 65\,\mic. At
60 and 100\,\mic, the coadded scans do show the expected square signal
where the source should be, but the deviation is not statistically
significant at the 3$\sigma$ level. Hence we use the 3$\sigma$
upper limits in our analysis. The photometric measurements are
compiled in Table~\ref{tab:photom}.

\subsection{Gemini/OSCIR observations}
\label{obs:gemini}
To constrain the size of the MIR emission from the source and obtain
photometric data outside the second silicate absorption band at
18\micron\ that was apparent in the ISOCAM spectrum (Paper~I), we
observed \sbs\ on the night of Dec\,9, 2000, at the Gemini-North
telescope, with the University of Florida mid-IR camera OSCIR. OSCIR
uses a 128$\times$128 pixel Si:As detector with a plate scale of 0.089
$\arcsec$/pixel providing a field of view of
11$\arcsec\times$11$\arcsec$ on the sky\footnote{Detailed information
on the instrument is available on the web at \\
http://www.gemini.edu/sciops/instruments/oscir/}. We used the N-wide
filter ($\lambda_{ref}$=10.8\,\mic\ and
$\Delta\lambda$=4.61\mic) and the Q3 filter ($\lambda_{ref}$=20.97\,\mic\ and
$\Delta\lambda$=1.05\,\mic)\footnote{These bandpasses are computed
assuming a $\lambda\,f_{\lambda}$=ct spectral convention. This
convention is thus identical for the ISOCAM, ISOPHOT and
Gemini/OSCIR data presented in this paper.}. In the rest of this
paper, we will refer to these filters as the 10.8 and 21\,\mic\
filters.

All the observations were performed using the standard technique of
chopping and nodding, with a chop throw of 15$\arcsec$ in
declination.  To obtain the most accurate photometry as possible, we
alternatively observed the source and standards at both wavelengths.
The standard stars used were $\beta$\,Peg, $\delta$\,Eri, and
$\alpha$\,Tau.  Flux density estimates for the standard stars were
calculated using the SED's published by \citet{Cohen99}.  From this,
it appeared that although the seeing remained constant during the
observations, at 0$\farcs$7 at 21\,\mic\, and 0$\farcs$43 at
10.8\,\mic, the sky transparency changed during the 10.8\,\mic\
observation of \sbs, after which it remained stable to within $\pm$5\%
for the rest of the observations.

Custom routines were used to stack the data appropriately to extract
the source signal, however since the source is faint even for
Gemini/OSCIR, no shift-and-add was possible.  During our observations
the OSCIR detector exhibited excess noise in one of its 16 output
channels. The main effect of this problem was an offset one, rather
than a gain one.  This is very reminiscent of the dark current problem
encountered on ISOCAM and, to remove this noise, we
applied the same ISOCAM algorithm to the OSCIR data to remove that
striping pattern \citep{Starck99}.  On these corrected images, the
source is clearly detected at both wavelengths.  On the reduced images,
the source appeared to be point-like or only slightly more extended
than the PSF. Therefore the data were ideally suited for filtering and
detection based on a wavelet decomposition of the image (compact
source in a large image with little or no background structure).  To
perform the detection and photometry of the object, we have used the
{\it MR1} package\footnote{See http://www.multiresolution.com}.  This
is a wavelet-based data reduction toolkit that implements all the
methods described in \citet{Starck98}.  The principle of the method is
to decompose the image in a cube where each plane holds only the
structures of a characteristic spatial scale, filter these planes,
apply a detection algorithm to identify significant deviations in the
planes, and reconstruct the detected objects.  In this process we are
helped by the fact that the PSF is extremely over-sampled and thus
even the smallest significant structures are on a larger scale than
most of the noise.  This process resulted in a clear detection of the
galaxy in both wavelengths with a respective positioning well inside
the relative pointing accuracy.  To constrain our global photometric
accuracy, we have performed simulations of the filtering and detection
process.  At the level of the source signal, with respect to the noise
level, the galaxy is detected in 100\% of the simulations.  However at
such faint levels, the photometric accuracy is poor, i.e. typically
30\% at both wavelength.  Taking into account the transparency
variations during the 10.8\,\mic\ observation, the resulting
photometric accuracy for that wavelength is 50\% (see
Table~\ref{tab:photom}).  One should note that with this wavelet
processing, it is not possible to define a S/N or a standard deviation
that could be tied to the source flux, as the noise is essentially
filtered out when we perform the photometric measurement.  Only
simulations of the detection process can indicate the validity of the
source.  The uncertainties we quote here therefore concern only the
photometric calibration of our data, and not the source detection.
Uncertainties attached to the source detection can be
estimated from the fraction of the simulations that either do not
detect the source or produce a false detection at the same flux level.
Our experiments show that this fraction is negligible.

In the resulting images, the source has a morphology similar to that
of a point source.  We therefore have no evidence for an extended
component to the MIR emission.  We note however that with a seeing
FWHM of 0$\farcs$43 at 10.8\,\mic\ we are not able to confirm or
contradict the conclusion of \citet{Dale01} that the infrared source
has a FWHM of 0$\farcs$31.

\section{Model of the infrared SED}
\label{model}
\subsection{Choice of a radiation transfer model}
\begin{figure}
\includegraphics*{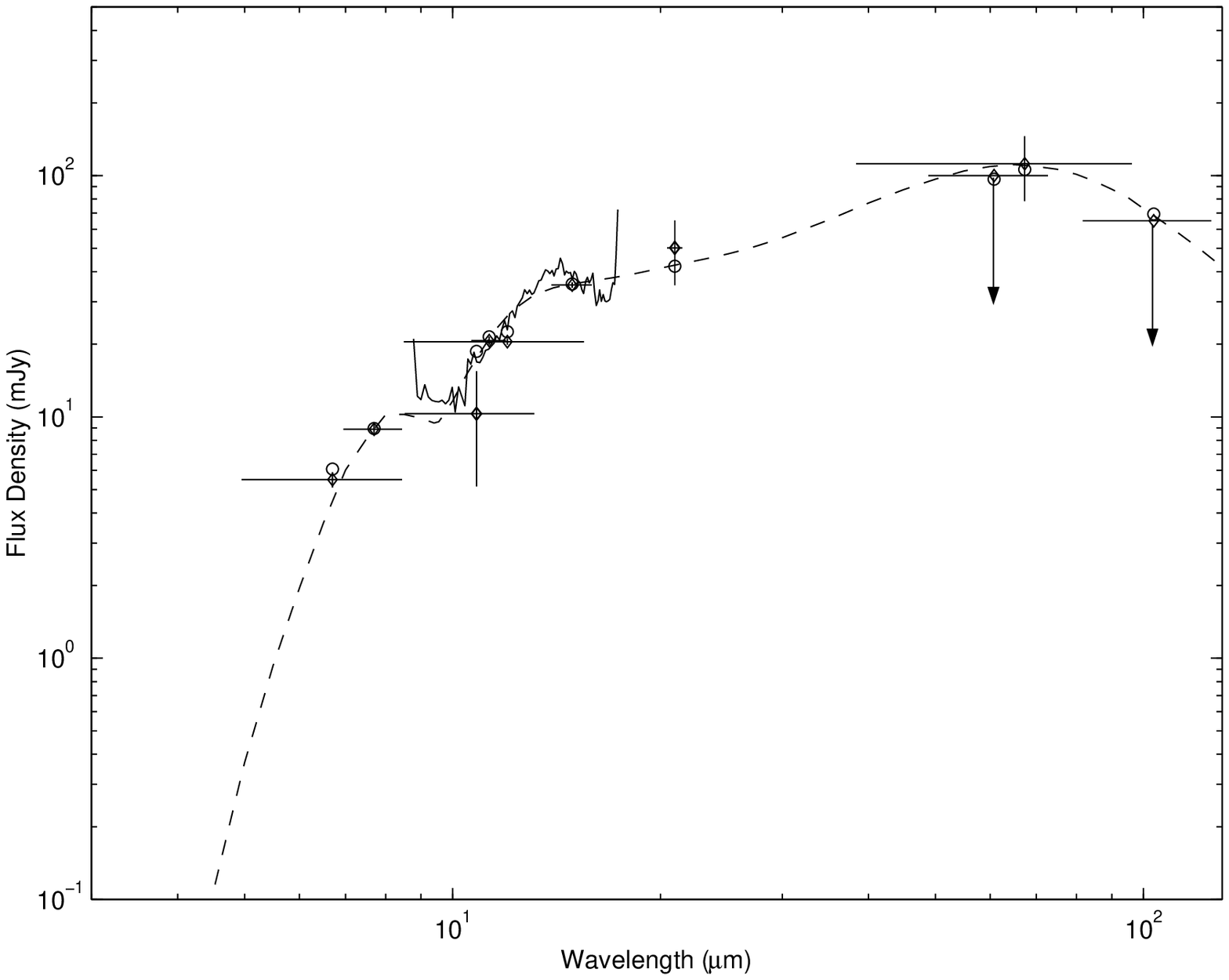}
\caption{Our best DUSTY model (broken line) superposed on the SED of \sbs\
(see Table~\ref{table:sbs_bestmodel} for its parameters). Symbols with
error bars represent the observed broadband photometric data, while
the solid line is the ISOCAM spectrum. Open circles represents model
fluxes in the same bandpasses, synthesized from the model.
}
\label{figure:sbs_bestmodel}
\end{figure}

The global infrared spectral energy distribution (SED) of \sbs\
is displayed in Figure~\ref{figure:sbs_bestmodel}.  It now shows a
further striking feature with respect to what was presented in
Paper~I, namely that the SED peaks at 60\,\mic, a much shorter
wavelength than what is observed in normal galaxies.  This places
\sbs\ in the category of galaxies called ``60\,\mic\ peakers'' in
the IRAS language.  Galaxies with this type of infrared SED are
either compact starburst or Seyfert with relatively high NIR
extinction \citep{Hei99}.  Silicate absorption in the MIR is not
uncommon, though it is rather restricted to the Seyfert galaxies
\citep{Lau00}. That the emission peaks at
60\,\mic\ probably indicates that \sbs\ lacks the dust phase that
is most common in other galaxies, the diffuse phase, exposed to
the diluted radiation of all the stars in the galaxy.  Rather, the
dust has to be quite close to the energy sources. For instance, if
we use the models of \citet{dbp90}, which assume an optically thin
line of sight from the radiation source to the grains, we can
compute the maximum distance at which grains have to be from a star cluster
to produce a ``60\,\mic\ peaker'' SED. Table~9 in \citet{dbp90} lists
the SED of grains exposed to the radiation of an O5 star as a
function of distance. One should be as close as
$\sim$2\,pc of such a star to observe a significant peak at
60\,\mic. For an SSC of 500 O5 stars, this converts to a distance
of $\sim$40\,pc. Given that optically visible SSCs have
core-haloes structure with characteristic sizes of 3 and 30\,pc
\citep{Oco94}, this forces the dust to be the closest
possible to the SSC.

This predominance of warm dust in the IR SED and the presence of a
silicate absorption band at 10\,\mic\ indicate that we can probably not assume
that the dust is optically thin to the heating radiation.  These
properties also imply that the dust will have a profound impact on the
spectral shape of the radiation from whatever source is heating it.
In this paper, we take advantage of the fact that we have a good
sampling of the infrared SED, as well as a very precise description of
the optical-UV SED, to constrain and model the transfer of radiation
from SSCs through the dust phase.

We used DUSTY \citep{Ive97,Ive99} to reproduce the SED of \sbs.
The currently available version of DUSTY takes into account
absorption, emission and scattering by dust.  By correctly treating
the radiation transfer process, it allows for the possibility that
colder dust absorbs radiation emitted by the hotter dust phases, i.e. dust
self-absorption, an effect that is neglected systematically when the
dust phase is simply treated as a screen (such as in Paper~I).  Its
two main limitations are (1) that it solves the problem of radiation
transport only in a spherical environment and (2) that it does not
include the treatment of stochastic heating\footnote{When the internal
energy of a grain becomes small compared to that of a single photon,
each absorption produces a spike in the grain temperature, followed by
cooling.  The grain never reaches thermal equilibrium and its
temperature history reflects the absorption of each photon. In a
given radiation field, it is always the smaller end of the size
distribution that will undergo stochastic heating.}.  We will
come back in section~\S\ref{discussion} on the consequences of these
limitations but we already note that (1) \sbs\ is located too far away
for us to be able to give prescriptions on the correct geometry for
the dust distribution, and (2) the SED shows no sign of UIB, which
implies that a much smaller fraction of the dust phase undergoes
impulsive heating than in more normal galaxies.

Finally we note that a second model exists that treats the same
problem in a more general way (i.e. the DIRTY model, \citet{Gord01,
Mis01}) using a Monte-Carlo approach, while DUSTY solves the problem
exactly. However this model is not yet in the public domain.

For DUSTY, we just have to specify the normalized spectrum of the
radiation source, i.e. the central star cluster, the dust composition
mix and its radial distribution, and the code calculates the dust
temperature radial distribution and the emerging radiation field.
Note that DUSTY uses the self-similarities included in the transfer
problem to simplify the computation, so all the output results are
dimensionless and have to be scaled back to the observed SED (see
section~\S\ref{results}).

\subsection{Input parameters for DUSTY}
As the input radiation, we used the spectrum from a 5 Myr old
starburst calculated by Starburst99 \citep{Lei_etal99} with a
[1-100]~\msolar\ mass range.  The
effect of the burst age (from 3 to 25 Myr) on the emerging SED
is negligible once the optical depth is sufficiently high, hence the
age of the central starburst is unconstrained by the fit. This
particular choice of burst age is motivated by the NIR analysis of
\citet{Van00}, and we come back to this in
section~\S\ref{subsect:star_cluster}. Note that the stellar mass
we deduce in this section is dependent not only on the age of the
cluster, but also on the stellar mass range. Inclusion of lower-mass
stars, though not noticeable in the SED and luminosity of the source,
leads to a higher total stellar mass.

Most of the free parameters of the model reside in the description
of the dust located around the source.  They are (1) the dust
chemical composition, (2) the temperature at the inner edge of the
dust shell, (3) the dust grain size distribution, (4) the
normalized density law along the shell radius, and (5) the optical
depth through the full dust cocoon.

\begin{figure}
\includegraphics*{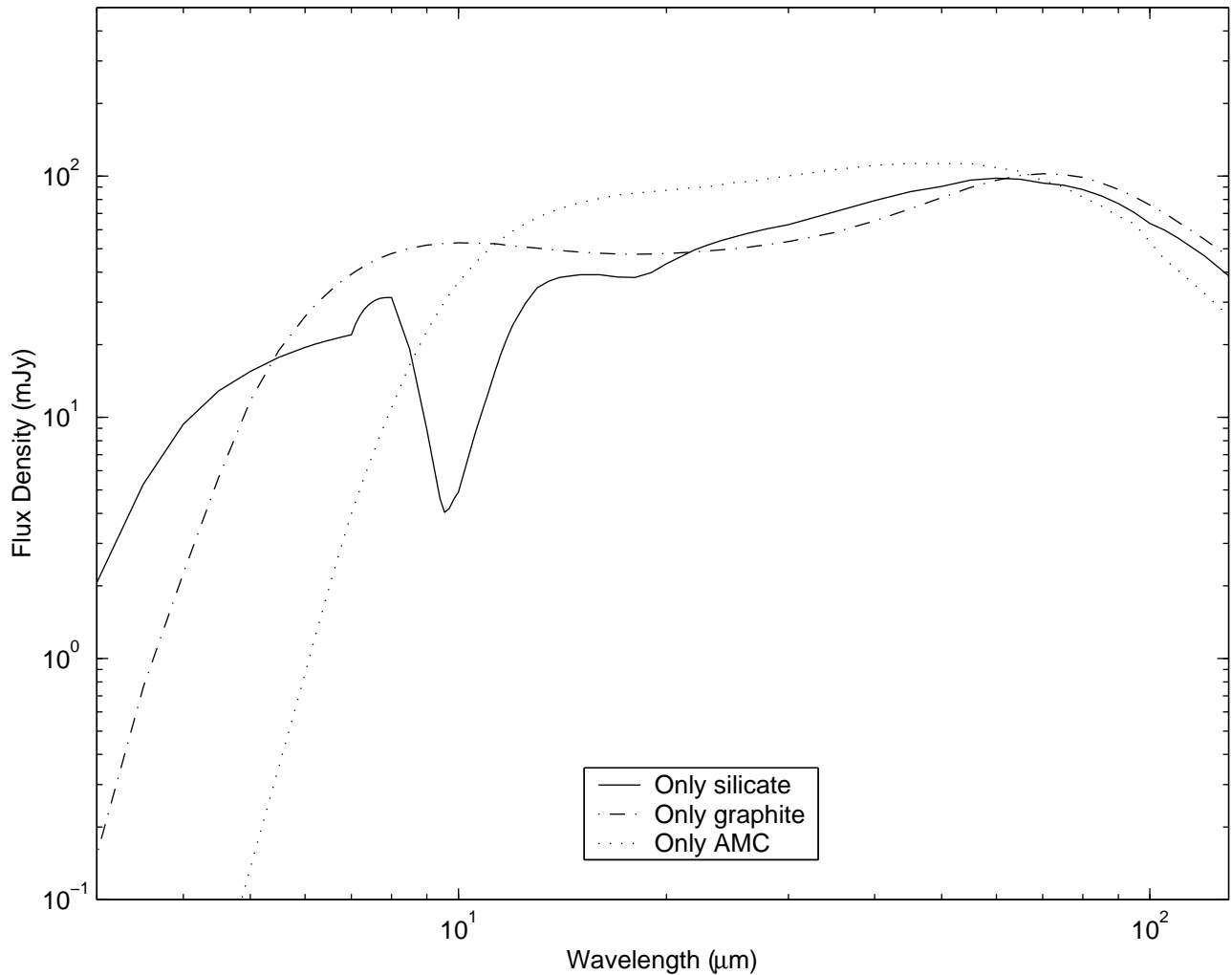}
\caption{Emerging SED for pure graphite (dash-dotted line),
silicates (solid line) and amorphous carbon (dotted line) dust
phases, exemplifying the impact of each dust specie. The other
parameters of the model are the same as those which were used to
reproduce the SED of \sbs\ (see Table~\ref{table:sbs_bestmodel}).
}
\label{figure:elements}
\end{figure}

The dust composition can be chosen from a variety of grain types, but
we decided to stick with the commonly used composition \citep{Wei01}:
silicate and graphite from \citet{Dra84}, and amorphous carbon from
\citet{Han88}.  The relative proportion of each of the component is a
free parameter.  The chemical mixture of the grains is more easily
constrained when we have a detailed spectrum, but we can still assess
the presence or absence of a grain type from the broad-band SED. To
exemplify how each dust component leaves its mark on the output SED,
Figure~\ref{figure:elements} shows the behavior of the SED for dust
composed of a single element.  Obviously the depth of the absorption
at 9.8\,$\micron$ is very sensitive to the relative proportion of
silicate.  In the absence of silicate, graphite will be responsible
for most of the emission below 9\,$\micron$, while amorphous carbon
will mostly fill the range between 20 and 100\,\mic.

The separation between the dust shell's inner face and the
radiation source is prescribed by the dust temperature at the
inner radius $T_1$.  This is in fact the only parameter of DUSTY
that has a dimension.  Thermal equilibrium of dust at the inner
radius links the temperature, the central cluster SED and the
inner radius.  Since the cluster SED is normalized, choosing $T_1$
selects the inner radius $r_{1}$.

The grain size distribution, based on an MRN-type \citep{Mat77}
distribution, $n(a) \propto a^{-q}$ for $a_{min} \leq a \leq
a_{max}$, is very critical, as it affects strongly the shape of
the SED: the larger number of smaller grains there are, the more flux we
observe in the 8\,\mic\ region.  The lower and upper cut-off have
the same practical effect as $q$.

\begin{figure}
\includegraphics*{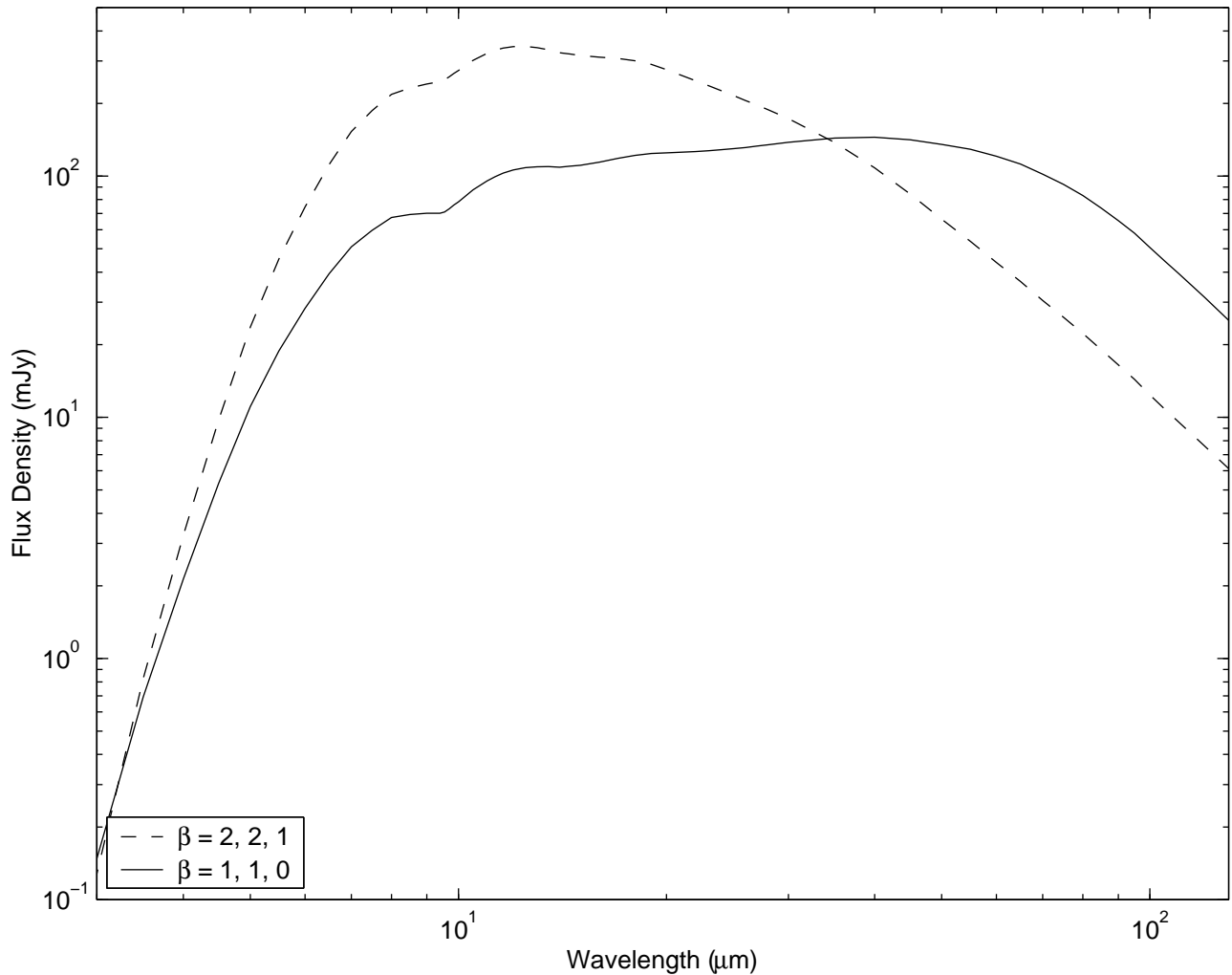}
\caption{Emerging SED for different values of $\beta$.  The
breaking points of the power-law distribution are at $r_1 = 10,
100$\ and 1000. The temperature (700K) and chemical composition of
the grains are the same for the two distributions as those which
were used to reproduce the SED of \sbs\ (see
Table~\ref{table:sbs_bestmodel}).
}
\label{figure:density}
\end{figure}

The dust distribution is spherical and has a radial density
dependence that we choose to follow a broken power-law $\eta
\propto y^{-\beta}$, where $y$ is the radial position normalized to
the inner radius of the dust shell $r_1$.  In DUSTY the dust
extends from $y=1$ to $y=1000$. The position of the breaking
points as well as $\beta$ are difficult to constrain, as we do not
know for sure the matter distribution in a SSC, nor that which
should be present in the cloud(s) where SSCs form. We separated
the shell in three zones, from $r_1$ to 10$r_1$, from 10$r_1$ to
100$r_1$ and from 100$r_1$ to 1000$r_1$, with each of them having
its own radial dependency $\beta$.  This separation is not
completely arbitrary.  It is first done because no acceptable fit
of the observed SED was possible with a single dust zone.  Then it
is introduced to allow some flexibility on the radial dependance
of the dust density, and also to understand the effects of the
radial density on the emerging spectrum.  In
Figure~\ref{figure:density} we show some of the effects that a change
in $\beta$ in the different zones has on the emerging SED.
For instance, with a rapid drop of density in the first
zone ($\beta = 2$ or 3), more dust can be heated to high
temperature, giving a rise in the flux at short MIR wavelengths. On
the opposite, flatter density profiles shift the SED toward
longer wavelengths as more dust is far away from the heating source.

Finally, the optical depth is the most important parameter, since it
critically determines how much dust is needed to produce
the observed SED.

To summarize, the free parameters in the model are the optical
depth $\tau$, the dust inner shell temperature $T_{1}$, the exponents
$\beta$ for the three density zones, the relative proportions of each
dust components, and the parameters of the dust size distribution,
$a_{min}$, $a_{max}$, and $q$.  This rather large number of parameters
is constrained by 9 broad-band values of flux and the ISOCAM spectrum, which
provides 25 additional independent measurements. Hence the
fit is over-constrained.

\section{Results from the model}
\label{results}
The best model fitting the SED of \sbs~is presented in Figure
\ref{figure:sbs_bestmodel} and the parameters used ($\tau$,
$T_{1}$, $\beta({\rm r})$, $a_{min}$, $a_{max}$, $q$ and dust
chemical composition) are given in Table
\ref{table:sbs_bestmodel}.  Comparison of a DUSTY SED with the
observed one is made by convolving the model SED with each
filter's bandpass.  We choose not to deredden the observed SED
from the effect of foreground Galactic dust as the amplitude of
this correction is negligible from the MIR upward \citep{Rie85}. A
$\chi$-square procedure is used to determine the best model. This
model reproduces all the photometric points we have obtained for
\sbs\ except one, which is our own very uncertain 10.8\,\mic\
Gemini measurement, and falls neatly within the ISOCAM spectrum
uncertainty.  The most salient result of this model is that the
SED requires a fairly high optical depth ($\tau = 30$), higher
than derived in Paper~I, and obviously much higher than
derived by \citet{Dale01} or \citet{Izo97}. We will come back to
that result in section~\S\ref{discussion}.

Exploration of the $chi$-square fit results allows for a
quantification of the range of acceptable values for the model
parameters. Acceptable fits are obtained when our parameters stay
within the following ranges around the nominal values listed in
table~\ref{table:sbs_bestmodel}: $\pm$5\% for the abundance fraction of
each dust component, $\pm$100~K for the internal temperature $T_{1}$,
$\pm$0.05\,\micron\ for $a_{min}$, and $\pm$2 on the optical depth. The
density distribution of the inner zone is very well constrained by
the observed SED as the wavelength range we sample is well adapted to
the temperature range in that region. The outer zone is less
constrained as we lack submillimeter data. However a steeper density
fall-off would not fit the SED. For the same reason, the upper size
limit of the grain distribution is not well constrained. These two
effects go in the same direction: the dust mass could be increased by
allowing larger grains or a shallower density profile in the outer
zone.

As we will see later, a critical information that can be deduced from
the fit is the size of the inner cavity where the radiation source of
DUSTY resides ($r_1$). The ranges quoted above allow for a 10\%
variation in the size of this cavity.

   From the best fit model, a number of important physical parameters can
be derived, which have their importance in the context of
star-formation in low-metallicity objects.  Among them the most
important ones are the bolometric luminosity of the enshrouded source,
and the total dust mass implied by the spectrum.  One must remember
that DUSTY is a scale-free modelling of the radiation transfer
problem, thus a number of arithmetic steps are needed to derive
absolute quantities such as a mass and a luminosity. Along these
steps the distance to \sbs\ will have to be used, which introduces
another source of uncertainty in all our deductions (see below).

\subsection{Parameters of the central starburst}
\label{subsect:result_sb}
The absolute bolometric luminosity of the central source is the
simplest parameter to derive from the model: the global scaling factor used to
match the DUSTY SED to the observed one allows to integrate the
complete IR-submm SED. This results in a central stellar
luminosity of 3.8$\times$10$^9$ \lsolar.  With the assumptions that the
central source is a 5\,Myr old starburst described by Starburst99,
this translates into 2$\times$10$^6$ \msolar~of stars. This compares
well with the value of 6.6$\times$10$^6$\msolar~obtained by
\citet{Hun01} based on Br$\alpha$ observations and an obscuration
of 15 visual magnitudes.  Since any acceptable fit requires a
relatively high optical depth, implying that the input
radiation is completely reprocessed by dust, there is a large
range of acceptable burst ages and IMFs for the central sources.
The only parameter of the central cluster which is well constrained
is its bolometric luminosity. One should note however that since
dust is more efficient in absorbing UV light, the emerging SED
also constrains the input SED, but to a lesser extent.

\subsection{Determination of the total dust mass}
\label{subsect:result_dust_mass}
Obtaining the total dust mass implied by the model is less
straightforward.  Basically we need to integrate the dust density over
the spherical shell.  This means recovering the actual physical
dimensions of $r$ and $\rho$ which are both normalized in DUSTY. First
we derive $r_{1}$, which is the inner radius of the shell.  $r_{1}$ is
related to the temperature $T_{1}$, and the absolute luminosity, both
of which are known.  In fact DUSTY provides a computation of $r_{1}$
for a 10$^{4}$\,\lsolar\ luminosity; a simple scaling to the actual
luminosity derived above provides $r_{1}=0.11$\,pc, with an
uncertainty of $\pm$10\% due to the range of models that provide an
acceptable fit to the SED (see above).  Since the shell
extends to 1000$r_{1}$ the physical dimension of the system is 110\,pc
(but note that the observable size will depend on the selected
wavelength).  We will come back to the meaning of these physical
dimensions, and in particular to their comparison with observed sizes
for the SSCs or globular clusters in other galaxies in
section~\S\ref{discussion}. The outer size derived above falls below the
spatial resolution in the Q-band
(diffraction-limited resolution of 173\,pc), and is slightly above that in
the N-band (diffraction-limited resolution of 90 pc).

Recovering the absolute value of the dust density is more complicated
because both the grain size distribution and the radial dependence
of the density are normalized. To derive these two normalization
factors, we use the fact that they are involved as well in the
determination of the optical depth and can be condensed both in
the $\tau$ and the dust mass equation into a single constant $C$. We
have the following relation between $\tau$ and $C$:
\begin{equation}
\tau = C\int_{r_1}^{r_{max}} \sum_{i} p(i) \int_{a_{min}}^{a_{max}}
Q_{eff}^i(a)\pi a^2 a^{-2.5} \rho(r) \ud r \ud a
\end{equation}
where $Q_{eff}^i$ is the effective scattering and absorbing
coefficient for dust component $i$ (here silicate, graphite and
amorphous carbon), $p(i)$ is the relative proportion of each component,
and $\rho$ is the density distribution.  When $C$ is known, we get the
dust mass by a simple integration over the density and the grain size
distribution. Using an optical depth of 30 leads to a total dust
mass of 1.5$\times$10$^5$ \msolar. Because
of the formal similarity of the dust mass and optical depth equations,
the dust mass is proportional to the optical depth as given by DUSTY.

\section{Discussion}
\label{discussion}

\subsection{Model-dependency of the results}
\label{subsect:modeldep}
Before drawing conclusions about the derived properties of the
embedded source in \sbs, it is worthwhile to mention how uncertain
and/or model-dependent these properties are.  The best constrained
parameter is the bolometric luminosity, as it is already well mapped
by our measurements.  Next comes the stellar mass, because this is
derived from the luminosity.  It requires an assumption on the age,
which is a priori difficult to make (though see
section~\S\ref{subsect:star_cluster}).  By analogy with the general
star formation process, one can reason that the age of embedded
sources should be smaller than the age of already visible ones.  This
places an upper limit at 5\,Myr.  The mass is then relatively well
constrained since it changes only by a factor of 2 for ages between 1
and 5\,Myr.

The major source of uncertainty or model-dependency is
introduced by the fact that DUSTY handles only dust in thermal
equilibrium.  This may impact the value of the inner radius of the
dust shell, as dust undergoing stochastic heating can reach higher
temperature in lower radiation density environments.
Given that the luminosity is well constrained, we can quantify the
scale on which stochastic heating is likely to play a part.
Following \citet{Des86} and \citet{Tra98}, we see that grains of
sizes larger than 0.1\,\mic\ will reach thermal equilibrium for values of the
inner radius $r_{1}$ up to 400\,pc, while grains larger than 0.01\,\mic\
would reach thermal equilibrium for values of $r_{1}$ up to 4\,pc.
We also find from \citet{Pug89} that PAHs would be destroyed by
the radiation for values of the inner radius of up to 11\,pc.
The conclusion of this is that, at least for the first few parsecs at
the base of the shell, solely on the basis of energy considerations,
we can exclude both the existence of small grains, and an important
contribution of stochastic heating to the thermodynamics of the
system.
We
therefore feel confident that the assumption of the whole
population of dust grains being at thermal equilibrium at the base of
the shell is correct, and that the actual value of $r_{1}$ is
also well-constrained (but see also section~\ref{subsect:star_cluster}).

The effect of assuming thermal equilibrium over the whole shell is
harder to estimate. As radiation propagates through the shell, it
is reddened and thus its ability to be absorbed by grains or
destroy them is lessened. The effect of allowing smaller-sized
grains which undergo thermal fluctuations is basically to have a
higher emission at short wavelengths per unit mass. It is thus
likely that a model allowing thermal fluctuations would require
less mass than DUSTY. Yet most of the mass is actually provided by
the colder dust producing the long-wavelength emission and for which
the assumption of thermal equilibrium is correct, therefore the
correction due to thermal fluctuations on the dust mass is likely
small.

To be able to more accurately represent the situation in \sbs,
a model would actually require not only that thermal fluctuations be
allowed but also that the size distribution change with radius.  No
such model is currently available.

\subsection{The embedded star cluster}
\label{subsect:star_cluster}
The stellar sources still embedded in their dust and molecular gas
cocoons cannot be described with precision from direct observations, as
they are invisible in the UV and NIR range.
The only way to get to their fundamental
parameters is again by looking at the SED produced by DUSTY with the
assumed starburst spectrum.  We used a 5 Myr population starburst, as
proposed by \citet{Van00}, with a Salpeter IMF between 1 and
100\msolar.  It should be noted that given the optical depth,
the origin of the NIR emission collected by \citet{Van00} is
probably composite.  Our model shows that only a fraction of
$<10^{-3}$ of the stellar luminosity emerges shortward of 3\,\mic.
However the extinction in the K-band is only
$\sim$3\,mag, which means that emission lines produced in the gas
surrounding the cluster could be observable.  The age deduced by
\citet{Van00} comes mainly from  broad-band colors, but, as
stated by the authors, these are
highly contaminated by nebular emission.  Therefore it is likely that
this age is indeed representative of the actual age of the stellar
cluster.  Furthermore, if what we observe is a still embedded
star-formation site, we can adopt the age measured in
the NIR, since it should represent an upper limit to the age of the
embedded sources.  The total stellar mass thus inferred,
2$\times$10$^{6}\,$\msolar, is only a lower limit (more mass is required for
an older cluster to achieve the same luminosity).  The bolometric
luminosity, at 3.8$\times$10$^{9}\,$\lsolar, is much better constrained, as
it comes directly from the SED. How does this source compare with
other embedded or optically visible SSCs, or with globular clusters?

Though little statistics exists on the properties of SSCs, the derived
mass of the \sbs\ stellar source is rather typical.  For instance,
\citet{Smi01} get masses of 0.5-1.2$\times$10$^{6}\,$\msolar\ for SSCs
in NGC\,1705, NGC\,1569 and M82.  The embedded SSC in the Antennae has
a mass of 1.6$\times$10$^{7}\,$\msolar\ \citep{gil00}, while
\citet{Men02} measure masses in the range
0.65-4.6$\times10^{6}$\,\msolar\ for five young visible clusters in
the same galaxy.  The mass we derive is also similar to that of the
brightest MIR cluster seen by \citet{Vac02} in He 2-10.

Luminosity-wise, the \sbs\ source is remarkably similar to the radio
super nebula in NGC\,5253 \citep{Gor01}, which
requires 0.8-1.2$\times10^9$\,\lsolar\ to produce its observed radio
flux. Therefore, even if such a luminosity is extreme for a single
SSC, it is not the only object of its kind.

The fact that we observe this source in a chemically young object as
well as in more evolved galaxies is very interesting as it points
toward a common phenomenon for violent star formation, regardless of
metallicity.

A further advantage of having performed a model of the radiation
transfer in the object is that we now have access to intrinsic
scale lengths of the cluster.  To follow the geometry used by
DUSTY, all of the stars have to be inside $r_{1} = 0.11$\,pc.
This is remarkably compact.
The compact size leads to a very large
stellar density of 3.8$\times10^{8}$\,\msolar\,pc$^{-3}$.  It is
difficult to compare that density with that of other SSCs as these
are rarely resolved. We note that it is much higher than that of
globular clusters (for instance the peak stellar density of M\,15
is 1.6$\times$10$^{6}$\,\msolar.pc$^{-3}$ \citep{Mey97}, while the mean
stellar density of M\,80 inside its core radius of 0.3\,pc is
3.3$\times10^{5}$\,\msolar\,pc$^{-3}$ \citep{Mad80}, approximately
identical to that of the young galactic center Arches cluster
\citep{Ser98}).  It
is impossible to know whether the very high stellar density we
derive is a problem as few models deal with the
formation of SSCs. We note however that such a compact cluster
would have an uncomfortably large virial velocity dispersion (of the
order of 300\,km.s$^{-1}$, much larger than any observed velocity
dispersions in SSCs, e.g. \cite{Men02}), leading to a very small
dynamical time (of the order of a few hundred years).

This situation can be relaxed if the radius of the volume actually
occupied by stars is allowed to grow to 1\,pc. Indeed, with this
value,
the stellar properties of the SSC would become absolutely
average compared to SSCs and globular clusters. As mentioned in 
section~\ref{subsect:modeldep}, this could be feasible by allowing 
thermal
fluctuations in the model. With this, the inner radius could be increased by
the fact that
temperatures higher than those derived from the thermal
equilibrium equation could occur further from the center. We note
however that it is the smaller sized grains that can run into that
regime, and that they are also thought to be less resilient to
destruction by radiation.

A more realistic possibility to allow the volume occupied by stars to grow
beyond the one attributed by DUSTY is to consider that, contrary to
the situation assumed by the model where dust and stars are
segregated, they are mixed in an inner region. Such a situation
cannot be computed with DUSTY, but it is likely that it would make
little difference in the principal outputs of the model. First, it is
true that by allowing the stellar cluster to expand, we lower the
mean radiation density in the inner region of the system. This may
lead to a decrease of the dust temperature, and thus of the MIR
emission, which can be compensated either by thermal fluctuations
mentioned above, or by allowing grains to get closer to the radiation
sources. Thus the hot dust required to produce the MIR emission in
the inner part of the system (so that it is later absorbed) can still
be present in this ``mixed" geometry. With an expanded cluster and
dust mixed with it, we also decrease to total optical depth. However,
with the density profile of the structure (see
table~\ref{table:sbs_bestmodel}), most of the dust, and thus of the
optical depth, is located in the outer parts of the structure, and
will not be impacted by the changes occurring in the inner
parts. This kind of geometry can unfortunately not be explored by
DUSTY since having an extended radiation source breaks one of the
requirements to exploit the self-similarities in the radiation
transfer problem. We however feel confident that such a situation
would keep most of the important outputs of the model.

To be complete, we should also state that the distance to \sbs\ is
involved in the determination of the stellar density to a power of
-2.5 to -2.3 (depending on the actual mass to light ratio of the
cluster stars). This is a strong dependency. However to obtain more
standard values of the stellar density would require unrealistically
large errors on the distance determination.

Finally we want to emphasize again that the dust cocoon, as we
model it, is extremely efficient at blocking the light of the
stars: less than 10$^{-3}$ of the bolometric luminosity emerges
shortward of 3\,\mic. With a high enough spatial resolving power, this
would lead to exactly the same situation as observed in He\,2-10
where the visible up to K-band morphologies are very similar, but
completely different from the MIR morphology \citep{Vac02}.

\subsection{The dust properties in \sbs}
\label{subsect:dust_properties}
The first interesting point of our experiments with DUSTY is that
a pure silicate dust phase is ruled out, contrary to what was
suggested in Paper~I. The dust chemical and size composition is
very well constrained by the Gemini Q-band observation and by the
ISOCAM spectra.  The strong continuum of the ISOCAM spectra as
well as the relative shallowness of the two silicate bands are
highly indicative of an important contribution to the emission by
graphite and amorphous carbon, both carbon-based (see Figure
\ref{figure:elements}) even though no PAH signatures are visible
in the spectrum. The absence of the PAH bands is thus not a strong
argument in favor of the absence of any carbon-based dust.

That the galaxy may be undergoing one of its first burst of star
formation is not an argument against carbon-based dust either since
carbon dust is rapidly formed in the ejecta of supernova
\citep{Tod01}, which may have already exploded in other regions of the
galaxy, as indicated by the shell-like structures observed by
\citet{Thu97}.

The size distribution that we observe is also very interesting as it
is quite different from the standard MRN distribution, both in the range
of grain sizes allowed and in the exponent of the distribution (see
Table~\ref{table:sbs_bestmodel}).  Yet this size distribution is not
completely new as it is very similar to that deduced from extinction
studies around the central engine of AGNs.  For instance,
\citet{Mai01} showed that the size distribution that explains best the
extinction observed toward AGNs is depleted in small grains, has
a rather high maximum size ($\sim$1\,\mic), and follows a power law
with an exponent of $-2.5$, as found for the best-fitting model for
\sbs.  This obviously does not imply that we have a mini-AGN at the
heart of \sbs, and in fact all spectroscopic data show that the source
is powered by star formation\footnote{From our modelled SED we
synthesized the fluxes required for the diagnostic diagram of
\citet{Laur00}.  These showed that \sbs\ is placed far away from the
AGN region of the diagram.} (see e.g. \citet{Van00}).  It is however
quite consistent with model-independent considerations on the
energetics of the heating source: indeed if we express the mean energy
density as a function of $y=r/r_{1}$, neglecting for the moment the
presence of the dust, we obtain that $\rho_{E}(y) =
2.1\times10^{5}\,y^{-2}$\,eV cm$^{-2}$.  This means that at the base
of the dust shell, the energy density is 4 orders of magnitude higher
than the value that would allow PAHs, {\em i.e.} the smallest dust
grains, to survive \citep{Pug89}.

The fact that the model tends to exclude small grains also indicates
that we are dealing with a very young star-forming region.  Indeed,
shocks such as those generated by supernovae are very efficient in
destroying the large grains to replenish the smallest sizes (see e.g.
\citet{Jon96}).  A dust size distribution biased toward the large
grains suggests on the contrary that most of the dust is still as it
was when large molecular clouds condensed to form a proto-SSC and that
no significant sweeping by SN shocks has yet occured (dust
sizes are known to grow in dense environments, see e.g.
\citet{Mai02}).

This exclusion of small grains, both on the basis of the model and on
that of the energy content of the source also brings further support
to the choice of DUSTY to model the radiative transfer. As indicated
in section~\ref{model}, DUSTY only treats grain in radiative
equilibrium. For any given radiation density level, the smallest dust
grains will not reach thermal equilibrium but rather undergo thermal
spiking, reaching temperatures much higher than the theoretical
equilibrium ones \citep{Tra98}. The size threshold to switch from
thermal equilibrium to thermal spiking is a decreasing function of
the energy density. Here we have both a very high energy density, at
least in the inner part of the dust shell, and a high minimum size
for the dust grain. These two properties go in the same direction
which is to limit the importance of thermal spiking with respect to
thermal equilibrium.

Regarding the radial dependence of the dust density, we note that
it is impossible to reproduce the full SED of \sbs\ {\em and} the
ISOCAM spectrum with a steeper distribution than what we used. This
would lead to an overproduction of luminosity in the 10\,\micron\
region. This emission can be compensated by a higher optical
depth, but this would causes a too strong silicate absorption band
at 9.7\,\micron. The last region of the density distribution, from
100$r_1$ to 1000$r_1$, is mainly affecting the longer wavelength
SED, as it contains mostly rather cold dust.  It affects strongly
the dust mass, being colder and distributed over a large region, as
it is the case for all galaxies.  The 60\,\micron\ PHT observation
constrains the external part of the distribution: too steep a
distribution does not produce enough luminosity at this
wavelength.

As an aside, our fitted SED predicts fluxes of 0.55\,$\mu$Jy and
0.03\,$\mu$Jy respectively at 450\,\micron\ and 850\,\micron, a range
of fluxes that are unfortunately not accessible to current
sub-millimeter telescopes.

\subsection{The optical depth and dust mass}
\label{subsect:tau}
The most important result of this paper is that we confirm, on a much
more solid basis, the high optical depth of the dust cloud measured in
Paper~I. The reason for the difference between the value of 21 V mag
quoted in Paper~I and the present result of $\sim$30 resides in the
fact that Paper~I used the unrealistic assumption of a screen of dust,
while here we account more correctly for the transfer of radiation
through the dust.  Our result is thus completely at odds with that of
\citet{Dale01} that argued for an optically thin source.  Let us
examine the reasons for this discrepancy.  The choice of a model that
is able to solve radiation transfer obviously does not bias the result
toward high optical depths, as DUSTY is perfectly able to reproduce
optically thin SEDs.  On the opposite, \citet{Dale01} follow a number
of questionable steps to arrive at the result that $A_{\rm V}\simeq
1$: they derive their optical depth using the link between the
optical depth and the surface brightness of a blackbody emission.
While this relation is exact, they still need to represent the SED as
a sum of blackbodies, and have a measure of the size of the emitting
region.  Rather than attempting to model the SED on a physical basis,
they simply fitted the SED with two blackbodies.  We note that
somewhat arbitrarily large error bars were added to the measurements
presented in Paper~I and that the resulting SED shows quite a poor
match with the ISOCAM spectrum of Paper~I. Regarding the size of the
emitting region, they have assumed that the emission is uniform over
the size of the object, as resolved by their observation (80$\pm$5\,pc
FWHM).  This is possibly the most questionable step of the reasoning
as the deduced value of $A_{\rm V}$ is proportional to this surface.
Our computation of the radiative transfer shows that even if the dust
shell extends up to 100\,pc, its FWHM {\em as observed in the N-Band}
is much more compact, probably less than 1\,pc.  Even if the inner radius of
the shell was increased by thermal fluctuation (see
section~\ref{subsect:star_cluster}), we do not expect that the
emission would have a constant surface brightness, but rather show a
core-halo structure that invalidates the use of the full size of the
emitting region in the computation of $A_{\rm V}$ by \citet{Dale01}.

The dust mass derived using DUSTY is of 1.5$\times10^{5}$\,\msolar.  This
is very close to the estimate of \citet{Hun01} using an independent
method, and inside the interval determined in Paper~I. Not
surprisingly it is much more that the estimate of \citet{Dale01},
because the incorrect assumptions on the SED and on the object size
propagate in the dust mass computation.  Although this is a small
intrinsic quantity with respect to the 9.5$\times10^{8}$\,\msolar\ of
H{\sc I} present around the galaxy \citep{Thu99a}, this is rather
large for such a metal-deficient object.  Does the star formation
history of \sbs\ allow for the formation of this dust phase?  In young
starbursts, most of the dust is provided by type-II supernovae
(SNe\,II).  \citet{Hir02} have studied the problem of dust formation
in low-metallicity environments, such as \sbs.  Their work shows that
the dust mass we measure can be accumulated in approximately
5$\times10^{7}$\,yr of continuous star formation at a rate of
1\,\msolar.yr$^{-1}$.  \citet{papaderos98} showed from broad-band colors that
the brightest visible clusters have ages in the range
0.1-3$\times10^{7}$\,yr, while another set of $\sim$10 fainter clusters
have ages in the range 3-10$\times10^{7}$yr.  Therefore the amount of dust
we measure is nearly consistent with the idea that the dust was
produced in the recent star-formation sites currently observed in the
visible.  We note that in the context of a continuous star-formation
scenario such as proposed by \citet{Leg00} for I\,Zw\,18, we would
have much more time to build up the dust phase.  However in that case
we would have (1) to take into account destruction processes that are
neglected in \citet{Hir02}, and (2) make sure that dust which has been
exposed to supernovae shock waves has enough time to coagulate so that
we observed the biased size-distribution required by the SED.

\subsection{The lifetime of an embedded SSC}
\label{subsec:lifetime}
The question of the lifetime of an embedded SSC such as the one in
\sbs\ is an important one, as it has a heavy bearing on our ability
to correctly account for all star formation activity in a galaxy.
Dust in the envelope of the SSC is subject to three main forces
from the SSC: radiation pressure and impact from stellar winds
will blow the dust out, while gravitation will attract it inward.
Following \citet{Wei01} and \cite{Ten98} we can express the sum of
the radiation pressure and impact from stellar winds as:
\begin{equation}
      P_{rad+wind}(a) = \frac{1}{4\pi r^{2}} \left[\frac{<Q_{pr}(a)>
L_{\star}}{c} +
      {\dot M}_{wind}v_{wind} \right],
\end{equation}
where $L_{\star}$ is the bolometric stellar luminosity, $<Q_{pr}(a)>$ is
the luminosity averaged radiation pressure efficiency on a grain of
size $a$, ${\dot M}_{wind}$ is the cluster's wind mass-loss rate,
$v_{wind}$ is the wind terminal velocity.  From Starburst99, we
obtain the kinetic luminosity (${\dot M}_{wind}v_{wind}^{2}/2$) of the
cluster, 5.2 $10^{6}$\,\lsolar. We can therefore express the ratio of the
radiation and kinetic pressure as:
\begin{equation}
         P_{rad}/P_{wind} = 9.7\, 10^{-7} v_{wind} <Q_{pr}(a)>.
\label{eq:radwind}
\end{equation}

Thus assuming a wind terminal velocity of 10$^{6}$\,m.s$^{-1}$, typical
of main-sequence O stars,
we obtain that the two terms are of the same order of magnitude at
the base of the dust shell where $<Q_{pr}(a)>$ is 1-2.  \citet{Wei01} showed
that under anisotropic radiation, photoelectron emission and
photodesorption of molecules from grains can act as an added
pressure, in some cases increasing the radiation pressure
efficiency by a factor of 20. In that case, radiation
effects will dominate. As a side note, let us mention
that the observed linewidths of only
160-170\,km.s$^{-1}$, as observed for instance by
  \cite{Tur01} toward the radio supernebula in NGC\,5253, are not
incompatible with the value chosen here for the wind terminal velocity.
Indeed, what is measured
is the outflow of ionized gas (through the velocity broadening of the
nebular Br$\gamma$ emission line), which is coupled with the molecular
and dusty envelope of the SSC
and hence slowed down by wind momentum conservation.

How then does radiation pressure compare to gravity? \citet{Lao93}
express the ratio of the acceleration produced by these two forces in
the optically thin case as:
\begin{equation}
\Gamma(a) = \frac{g_{rad}}{g_{grav}} = \frac{3
L_{\star}<Q_{pr}(a)>}{16 \pi c a \rho G
M_{\star}},
\end{equation}
where $M_{\star}$ is the mass of the stellar cluster, and $\rho$
the density of the dust grain.  This optically thin case applies
well to the base of the dust shell and, using
$\rho$\,=\,3\,10$^{3}$\,kg.m$^{-3}$, this gives numerically
$\Gamma(a) = 360\,<Q_{pr}(a)>/a_{\mic}$.  We thus see that
radiation pressure overtakes gravity by several orders of
magnitude at the base of the shell, i.e. that it can be moving the
shell outward.  At the outer boundary of the shell the situation
may be different since there, the luminosity is heavily reddened
and therefore the luminosity-averaged radiation pressure efficiency will be
much lower.  Using the formula for $\Gamma(a)$ given by
\citet{Lao93} in the optically thick case, we see that in the
outer parts of the shell, gravity and radiation pressure almost
balance each other ($\Gamma(a)\simeq1.7$).  It is unlikely that
the effects described in \cite{Wei01} will occur at the outer
boundary of the shell since the radiation is now too soft to allow
photoelectron emission or photodesorption of molecules to occur.

Since this neglects the influence of an outer pressure, we have here
an indication that an embedded SSC may be a long-lived structure,
lasting at least till the first supernovae explode.  This is however
only indicative.  Indeed, a simple calculation of the motion of a
grain subject to radiation pressure, stellar winds and gravity from an
SSC such as the one we deduce in \sbs\ shows that it can be quite
efficiently blown away from the SSC. Timescales derived from such a
calculation are however incorrect since grains will feel a strong drag
from collisions with the gas that is associated to them.  This drag
will probably be even stronger at the base of the shell where
ionization of gas and grains can lead to a very efficient coupling of
the two phases \citep{Fer01}.  A detailed modelling of the dynamics of
such a region would therefore be a worthwhile endeavor in order to
constrain the lifetime of the SSC embedded stage.  We note that the
presence of similar objects in a number of sources (e.g. He\,2-10,
NGC\,5253, NGC\,4038/9) indicates that these lifetimes cannot be
extremely short.  On the other hand, the fact that the age of optically visible
clusters are in the range of a few Myr probably sets an upper limit to the
embedded stage lifetime.

\subsection{Implications for star-formation at high redshift}
\label{subsect:highZ}
\sbs, with an extremely low metallicity, is already a galaxy where
much of the star-formation activity is completely hidden from view
from UV to NIR, possibly for a significant fraction of the starburst
lifetime.  The 3.8$\times10^{9}$\,\lsolar\ bolometric luminosity of
the cluster translates into 3.8$\times10^{4}$ equivalent O7 stars.
This is a factor of 10 more than the number of equivalent O7 stars
required to power the visible clusters \citep{Hun01}.  This
demonstrates that even in extremely low-metallicity objects, the
visible-UV part of the SED may be relatively insignificant and is not
a reliable indicator of to the actual star-formation activity.  In
this respect, \sbs\ can be considered as a smaller and nearer analog
to the $z=2.56$ submillimer galaxy SMM\,J14011$+$0252 studied by
\citet{Ivi02}.  Therefore it could very well be that most of the
star-formation episodes that occur during the first phases of galaxy
formation are completely hidden from view in the short wavelength
part of the electromagnetic spectrum.  We already see that the
embedded SSC phenomenon happens in numerous dwarf galaxies,
irrespective of their metallicity.  Therefore it only remains to be
seen whether the process of star formation, as we observe it in dwarf
galaxies, provides a
suitable analog to the situation occurring in primeval objects.

\section{Conclusion}
We have modelled the infrared SED of the blue compact dwarf galaxy
\sbs\ with DUSTY, which solves consistently the radiation transfer in
a spherical distribution of dust.  From this modelling, we deduce that
\sbs\ harbors a deeply embedded super-star cluster, effectively hidden
under about 30 mag of visual extinction.  The low-metallicity of the
galactic gas did not preclude the formation of the 10$^5$ \msolar\ of
dust necessary to completely hide from the optical view the SSC. With
2$\times10^6$ \msolar\ of stars and an age of probably less than 5
Myr, the SSC has not been able yet to pierce through the cocoon of
dust and gas from which it formed, but it had a profound effect on the
dust size distribution: the hardness of the radiation
destroyed the smallest dust particles and the PAH but shocks did not
alter yet the larger size grains up to 1\,\micron.  The standard MRN
distribution, normally observed in quiescent galactic environment,
cannot reproduce the IR data we have at hand.  Instead, the dust
in the SSC environment is reminiscent of what we observe around AGN,
emphasizing the role of density and radiation hardness on the dust
grain size distribution.  If dust-enshrouded SSCs are commonly associated to
starbursting environment, the star-formation rate deduced by looking
at the rest-frame optical or UV should be taken with caution.  Even
with IR or MIR information care should be taken to use a correct
radiation transfer treatment and one should use an extinction law
suited to the radiation and gas density of the observed source.

\acknowledgements We warmly thank Martin Haas of the ISOPHT data
center in Heidelberg and Scott Fisher of Gemini North for assistance
and advice in the data reduction.  Jean-Luc Starck's guidance on the
use of wavelet analysis was deeply appreciated.  Many thanks to
Carmelle Robert, Jean-Ren\'e Roy and Pierre-Alain Duc for helpfull
discussions and comments on the manuscript. We also acknowledge the
critical reading of our anonymous referee that led us to a more
in-depth analysis of important parts of this paper, resulting in, we
hope, a more convincing study.

\clearpage

\clearpage

\clearpage

\clearpage

\begin{deluxetable}{llrr}
\tablewidth{0pt}
\tablecaption{Photometric data used in this paper}
\startdata
\hline \hline
$\lambda_{ref}$ & Instrument & Flux & 1$\sigma$ \\
\mic\           &            & mJy  & mJy       \\
\hline
10.8 & Gemini/OSCIR & 10.3   & 50\%$^{a}$ \\
21   & Gemini/OSCIR & 50.2   & 30\%$^{a}$ \\
60   & ISOPHOT      & $<$100 &            \\
65   & ISOPHOT      & 112    & 21         \\
100  & ISOPHOT      & $<$65  &            \\
\hline
\enddata
\tablenotetext{a}{These are photometric accuracies (see
section~\ref{obs:gemini})
}
\label{tab:photom}
\end{deluxetable}

\clearpage

\begin{deluxetable}{l|c|l}
\tablewidth{0pt}
\tablecaption{Best parameters for DUSTY to
reproduce the SED of \sbs}
\startdata
\hline
\hline
Optical Depth at 0.55 \micron &                        & 30.0 \\
\hline
Temperature                   & (K)                    & 700  \\
\hline
                                  & $r_1$ to 10$r_1$       & 1 \\
Density Law ($\beta$)         & 10$r_1$ to 100$r_1$    & 1 \\
                                  & 100$r_1$ to 1000$r_1$  & 0 \\
\hline
                                  & Silicate               & 23\% \\
Elements                      & Graphite               & 43\% \\
                                  & AMC                    & 34\% \\
\hline
                                  & Exponent               & -2.5 \\
Size distribution             & Lower cut-off (\micron)& 0.022 \\
                                  & Upper Cut-off (\micron)& 1 \\
\hline
\hline
\enddata
\label{table:sbs_bestmodel}
\end{deluxetable}


\begin{thebibliography}{}
\bibitem[Allamandola et al.(1985)]{All85} Allamandola, L.J., Tielens,
A.G.G.M., \& Barker, J.R. 1985, \apj, 290, L25

\bibitem[Calzetti et al.(1994)]{Cal94} Calzetti, D., Kinney, A. L., \&
Storchi-Bergmann, T. 1994, \apj, 429, 582

\bibitem[Cannon et al.(2002)]{Can02} Cannon, J. M., Skillman, E. D.,
Garnett, D. R. \& Dufour, R. J. 2002, \apj, 565, 931

\bibitem[Cohen et al.(1999)]{Cohen99} Cohen, M., Walker, R. G.,
Carter, B., Hammersley, P., Kidger, M., \& Noguchi, K. 1999, \aj,
117, 1864

\bibitem[Contursi et al.(2000)]{Con00} Contursi, A., Lequeux, J.,
Cesarsky, D., Boulanger, F., Rubio, M., Hanus, M., Sauvage, M., Tran,
D., Bosma, A., Madden, S., \& Vigroux, L. 2000, \aap, 362, 310

\bibitem[Dale et al.(2001)]{Dale01} Dale, D. A., Helou, G.,
Neugebauer, G., Soifer, B. T., Frayer, D. T. \& Condon, J. J.
2001, \aj, 122, 1736

\bibitem[D\'esert et al.(1990)]{dbp90} D\'esert, F. X., Boulanger, F.
\& Puget, J. L. 1990, \aap, 237, 215

\bibitem[D\'esert et al.(1986)]{Des86} D\'esert, F. X., Boulanger,
F., \& Shore, S. 1988, \aap, 160 285

\bibitem[Doublier et al.(2001)]{Doublier01} Doublier, V., Caulet, A.,
\& Comte, G. 2001, \aap, 367, 33

\bibitem[Draine \& Lee(1984)]{Dra84} Draine, B.T., \& Lee, H.M. 1984,
\apj, 285, 89

\bibitem[Ellis(1998)]{Ell98} Ellis, R. 1998, Nature, 395, A3

\bibitem[Ferland(2001)]{Fer01} Ferland, G. J. 2001, \pasp, 113, 41

\bibitem[Gilbert et al.(2000)]{gil00} Gilbert, A. M., Graham,
J. R., McLean, I. S., Becklin, E. E., Figer, D. F., Larkin, J. E.,
Levenson, N. A., Teplitz, H. I. \& Wilcox, M. K. 2000, \apjl, 533,
L57

\bibitem[Gordon et al.(2001)]{Gord01} Gordon, K. D.,
Misselt, K. A., Witt, A. N. \& Clayton, G. C. 2001, \apj, 551, 269

\bibitem[Gorjian et al.(2001)]{Gor01} Gorjian, V., Turner, J.L., \&
Beck, S.C. 2001, \apj, 554, L29

\bibitem[Hanner(1988)]{Han88} Hanner, M.S. 1988, NASA Conf.  Pub.,
3004, 22

\bibitem[Hartmann et al.(1988)]{Har88} Hartmann, L.W., Huchra, J.P.,
Geller, M.J., O'Brien, P., \& Wilson, R. 1988, \apj, 326, 101

\bibitem[Heisler \& de Robertis(1999)]{Hei99} Heisler, C. A., \& de
Robertis, M. M. 1999, \aj, 118, 2038

\bibitem[Hirashita et al.(2002)]{Hir02} Hirashita, H., Hunt, L. K.,
\& Ferrara, A. 2002, \mnras, 330, L19

\bibitem[Hunt, Vanzi, \& Thuan(2001)]{Hun01} Hunt, L.K., Vanzi, L., \&
Thuan, T.X. 2001,\aap, 377, 66

\bibitem[Ivezi\'c \& Elitzur(1997)]{Ive97} Ivezi\'c, \v Z., Elitzur, M.
1997, \mnras, 287, 799

\bibitem[Ivezi\'c et al.(1999)]{Ive99} Ivezi\'c, \v Z., Nenkova, M.,
\& Elitzur, M. 1999, User Manual for DUSTY, University of Kentucky
Internal Report, {\tt http://www.pa.uky.edu/~moshe/dusty/}

\bibitem[Ivison et al.(2001)]{Ivi02} Ivison, R. J., Smail, I.,
Frayer, D. T., Kneib, J.-P., Blain, A. W. 2001, \apj, 561, L45

\bibitem[Izotov et al.(1997)]{Izo97} Izotov, Y., Lipovetsky, V. A.,
Chaffee, F. H., Foltz, C. B., Guseva, N.G., \& Kniazev, A.Y. 1997,
\apj, 476, 698

\bibitem[Johnson et al.(2000)]{Joh00} Johnson, K.E., Leitherer, C.,
Vacca, W.D., \& Conti, P.S. 2000, \aj, 120, 1273

\bibitem[Jones et al.(1990)]{Jon90} Jones, A.P., Duley, W.W., \&
Williams, D.A. 1990, \qjras, 31, 567

\bibitem[Jones et al.(1996)]{Jon96} Jones, A.P., Thielens, A.G.G.M.,
\& Hollenback, D.J. 1996, \apj, 769, 740

\bibitem[Jones \& d'Hendecourt(2000)]{Jon00} Jones, A. P.,
d'Hendecourt, L. 2000, \aap, 355, 1191

\bibitem[King (1966)]{Kin66} King, I.R. 1966, \aj, 71, 64

\bibitem[Kobulnicky \& Johnson(1999)]{Kob99} Kobulnicky, H.A., \&
Johnson, K.E. 1999, \apj, 527, 154

\bibitem[Kunth \& \"Ostlin(2000)]{Kun00} Kunth, D., \& \"Ostlin, G.
2000, \aapr, 10, 1

\bibitem[Laor \& Draine(1993)]{Lao93} Laor, A., \& Draine, B. T.
1993, \apj, 402, 441

\bibitem[Laureijs et al.(2000a)]{Laureijs00} Laureijs, R. J., Klaas,
U., Richards, P. J., Schultz, B., \& \'Abrah\'am, P. 2000a, The
ISO Handbook, Vol.  V. SAI-99-069/Dc, Version 1.1

\bibitem[Laureijs et al.(2000b)]{Lau00} Laureijs, R. J., Watson, D.,
Metcalfe, L., McBreen, B., O'Halloran, B., et al. 2000b, \aap, 359, 900

\bibitem[Laurent et al.(2000)]{Laur00} Laurent, O., Mirabel,
I. F., Charmandaris, V., Gallais, P., Madden, S. C., Sauvage, M.,
Vigroux, L. \& Cesarsky, C. 2000, \aap, 359, 887

\bibitem[L\'eger \& Puget(1984)]{Leg84} L\'eger, A., \& Puget, J.L.
1984, \aap, 137, L5

\bibitem[Legrand et al.(2000)]{Leg00} Legrand, F., Kunth, D., Roy,
J.-R., Mas-Hesse, J. M., \& Walsh, J. R. 2000, \aap, 355, 891

\bibitem[Leitherer(1999)]{Lei99} Leitherer, C. 1999, in `Chemical
evolution from zero to high redshift', Eds.  Walsh, J., Rosa M.,
Lecture notes in Physics (Berlin: Springer)

\bibitem[Leitherer et al.(1999)]{Lei_etal99} Leitherer, C., Schaerer,
D, Goldader, J.D., Delgado, R.M.G., Robert, C., Kune, D.F., de Mello,
D.F., Devost, D. \& Heckman, T.M. 1999, \apjs, 123, 3

\bibitem[Lemke et al.(1996)]{Lemke96} Lemke, D., Klaas, U., Abolins,
J., \'Abrah\'am, P., Acosta-Pulido, J. et al. 1996, \aap, 315, L64

\bibitem[Madore(1980)]{Mad80} Madore B., 1980 in `Globular clusters',
Hanes, D., Madore, B (eds.) Cambridge University Press, p21

\bibitem[Maiolino(2002)]{Mai02} Maiolino, R. 2002, in `The
evolution of galaxies. II. Basic building blocks', M. Sauvage et al.
(eds.) in press

\bibitem[Maiolino et al.(2001)]{Mai01} Maiolino, R., Marconi, A., \&
Oliva, E. 2001, \aap, 365, 37

\bibitem[Mathis et al.(1977)]{Mat77} Mathis, J.S., Rumpl, W., \&
Nordsieck, K.H. 1977, \apj , 217, 425

\bibitem[Mengel et al.(2002)]{Men02} Mengel, S., Lehnert, M. D.,
Thatte, N., Genzel, R. 2002, \aap, 383, 137

\bibitem[Meylan et al.(1997)]{Mey97} Meylan, G., \& Heggie, D. C. 1997,
A\&AR, 8, 1

\bibitem[Meurer et al.(1995)]{Meu95} Meurer, G., Heckman, T.M.,
Leitherer, C., Kinney, A., Robert, C., \& Garnett, D.R. 1995, \aj, 110,
2665

\bibitem[Mirabel et al.(1998)]{mvc98} Mirabel, F., Vigroux, L.,
Charmandaris, V., Sauvage, M., Gallais, P., Cesarsky, C., Madden,
S. \& Duc, P. A. 1998, \aap, 333, L1

\bibitem[O'Connell et al.(1994)]{Oco94} O'Connell, R.W., Gallagher,
J. S., \& Hunter, D. A. 1994, \apj, 433, 65

\bibitem[Misselt et al.(2001)]{Mis01} Misselt, K. A.,
Gordon, K. D., Clayton, G. C. \& Wolff, M. J. 2001, \apj, 551, 277

\bibitem[\"Ostlin(2000)]{Ost00} \"Ostlin, G. 2000, \apj, 235, L99

\bibitem[\"Ostlin \& Kunth(2001)]{Ost01} \"Ostlin, G., \& Kunth, D.
2001, \aap, 371, 429

\bibitem[Papaderos et al.(1998)]{papaderos98} Papaderos, P., Izotov,
Y. I., Fricke, K. J., Thuan, T. X., \& Guseva, N. G. 1998, \aap, 338,
43.

\bibitem[Pisano et al.(2001)]{Pis01} Pisano, D.J., Kobulnicky, H.A.,
Guzman, R., Gallego, J., \& Bershady, M.A. 2001, \aj, 122, 1194

\bibitem[Puget \& L\'eger(1989)]{Pug89} Puget, J. L., \& L\'eger, A.
1989, \araa, 27, 161

\bibitem[Rieke \& Lebofsky(1985) ]{Rie85} Rieke, G.H., \& Lebofsky,
M.J. 1985, \apj, 288, 618

\bibitem[Sanders \& Mirabel(1996)]{San96} Sanders, D. B., \& Mirabel, I.
F. 1996, \araa, 34, 749

\bibitem[Sauvage et al.(1997)]{Stl97} Sauvage, M., Thuan, T. X. \&
Lagage, P. O. 1997, \aap, 325, 98

\bibitem[Sauvage et al.(2002)]{Sau02} Sauvage, M., et al. 2002, in
preparation.

\bibitem[Serabyn et al.(1998)]{Ser98} Serabyn, E., Shupe, D. \&
Figer, D. F. 1998, Nature, 394, 448

\bibitem[Silk \& Devriendt(2001)]{Sil01} Silk, J., \& Devriendt, J.
2001 in {\em The Extragalactic Infrared Background and its
cosmological implications}, IAU Symposium, Vol. 204, M. Harwit \& M.
G. Hauser, eds. ASP, 423

\bibitem[Smith \& Gallagher(2001)]{Smi01} Smith, L. J., \& Gallagher,
J. S. III 2001, \mnras, 326, 1027

\bibitem[Starck et al.(1998)]{Starck98} Starck, J. L., Murtagh, F., \&
Bijaoui, A. 1998, `Image processing and data analysis: the multiscale
approach', Cambridge University Press, Cambridge (GB)

\bibitem[Starck et al.(1999)]{Starck99} Starck, J. L., Abergel, A.,
Aussel, H., Sauvage, M., Gastaud, R., Claret, A., D\'esert, F. X.,
Delattre, C., \& Pantin, E. 1999, \aaps, 134, 135

\bibitem[Tenorio-Tagle \& Medina-Tanco(1998)]{Ten98} Tenorio-Tagle,
G., \& Medina-Tanco, G. A. 1998, \apj, 503, L171

\bibitem[Todini \& Ferrara(2001)]{Tod01} Todini, P., \& Ferrara,
A. 2001, \mnras, 325, 726

\bibitem[Thuan et al.(1999a)]{Thu99a} Thuan, T. X., Lipovetsky, V. A.,
Martin, J. M., \& Pustilnik, S. A. 1999a, \aaps, 139, 1

\bibitem[Thuan et al.(1999b)]{Thu99} Thuan, T.X., Sauvage, M., \&
Madden, S. 1999b, \apj, 516, 783

\bibitem[Thuan et al.(1997)]{Thu97} Thuan, T.X., Izotov, Y.I., \&
Lipovetsky, V.A. 1997, \apj, 477, 661

\bibitem[Tran(1998)]{Tra98} Tran, Q. D. 1998, PhD Thesis, Universit\'e
Paris XI

\bibitem[Turner et al.(2000)]{Tur00} Turner, J.L., Beck, S.C., \& Ho,
P.T.P. 2000, \apj, 532, L109

\bibitem[Turner et al.(2001)]{Tur01} Turner, J. L., Crosthwaite, L.
P., Meier, D. S., Beck, S. C. 2001, AAS 198, \# 09.05

\bibitem[Vacca et al.(2002)]{Vac02} Vacca, W. D., Johnson, K. E.
\& Conti, P. S. 2002, \aj, 123, 772

\bibitem[Vanzi et al.(2000)]{Van00} Vanzi, L., Hunt, L.K., Thuan,
T.X., \& Izotov, Y.I. 2000, \aap, 363, 493

\bibitem[Weingartner \& Draine(2001)]{Wei01} Weingartner, J.C., \&
Draine, B.T. 2001, \apj, 548, 296

\end{thebibliography}
\end{document}